\documentclass[prl,aps,twocolumn,showpacs]{revtex4}
\usepackage{latexsym}
\usepackage{color,graphicx,shortvrb}

\begin{document}
\title{Microscopic Dynamics in a Strongly Interacting Bose-Einstein Condensate}
\author{N.~R. Claussen, E.~A. Donley, S.~T. Thompson, and C.~E. Wieman}
\affiliation{JILA,
  National Institute of Standards and Technology and the University of
  Colorado, and the
  Department of Physics, University of Colorado, Boulder, Colorado
  80309-0440}
\date{\today}

\begin{abstract}

An initially stable $^{85}$Rb Bose-Einstein condensate (BEC) was
subjected to a carefully controlled magnetic field pulse in the
vicinity of a Feshbach resonance.  This pulse probed the strongly
interacting regime for the condensate, with calculated values for
the diluteness parameter ($na^3$) ranging from 0.01 to 0.5.  The
field pulse was observed to cause loss of atoms from the
condensate on remarkably short time scales ($\geq$10$~\mu$s). The
dependence of this loss on magnetic field pulse shape and
amplitude was measured.  For triangular pulses shorter than 1~ms,
decreasing the pulse length actually {\it increased} the loss,
until extremely short time scales (a few tens of microseconds)
were reached.  Such time scales and dependencies are very
different from those expected in traditional condensate inelastic
loss processes, suggesting the presence of new microscopic BEC
physics.
\end{abstract}


\maketitle

The self-interaction strength that determines most of the
properties of a Bose-Einstein condensate (BEC) is characterized
simply by the s-wave scattering length, $a$ \cite{review}.  Near a
Feshbach resonance \cite{Tiesinga1993a}, the scattering length
depends strongly on the magnetic field.  This property has been
used to vary the self-interaction energy in BEC \cite{Inouye1998,
Stenger1999, Cornish2000, Roberts2001, Donley2001}.  In
particular, we have used the Feshbach resonance to create stable
$^{85}$Rb condensates \cite{Cornish2000}.  We also varied the
scattering length and studied the effect on the condensate
--- most notably its collapse when the scattering length was made
negative \cite{Roberts2001, Donley2001}.

Here we discuss BEC behavior in the positive $a$ region near the
$^{85}$Rb Feshbach resonance, where the self-interaction is large
and repulsive.  We used rapid magnetic field variations to probe
the strongly interacting regime in the condensate and to
investigate the possibility of collisional coupling between pairs
of free atoms and bound molecules
\cite{Timmermans1999,Drummond1998,Abeelen1999,Yurovsky1999,Mies2000,Holland2001}.
We changed the magnetic field to approach the Feshbach resonance
from above, which causes the positive scattering length to
increase.  As $na^3$ becomes comparable to or larger than one, the
customary mean-field description of the dilute gas BEC, which
assumes no correlations between the atoms, breaks down.
Correlations between the atoms become increasingly important and a
BEC with sufficiently large $na^3$ will eventually reach a highly
correlated state such as that found in a liquid. We are interested
in the nature of this transition to a highly correlated state and
the time scale for the formation of the condensate correlations.
In addition to such interesting physics, our time dependent
experiments also allow us to probe the coupling between atomic and
molecular BEC states that are degenerate at the Feshbach
resonance.  Several authors have proposed mechanisms for
transferring atom pairs into molecules using time-dependent
magnetic fields near resonance
\cite{Timmermans1999,Drummond1998,Abeelen1999,Yurovsky1999,Mies2000,Holland2001}.
To investigate such physics we examined the response of the
condensate when we briefly approached the resonance by applying a
short magnetic field pulse (much shorter than the period for
collective excitations). The most obvious feature of the BEC
response was a loss of atoms that increased when we made our
pulses shorter, until the loss became very large for very short
pulses.

Bose-Einstein condensates always disappear if one waits long
enough.  The dominant loss process was found to be three-body
recombination into molecules \cite{Burt1997,myatt_thesis}, with a
time dependence characterized by a simple rate equation and rate
constant.  All previous observations of condensates are consistent
with a mean-field description that includes such a
density-dependent loss process. In the vicinity of a Feshbach
resonance the decay rates have been seen to increase dramatically
\cite{Stenger1999,Cornish2000,jakethesis}, becoming so large that
various novel coherent conversion processes have been put forth to
explain their magnitude \cite{Abeelen1999,Yurovsky1999,Mies2000}.
However, the observations in
Refs.~\cite{Stenger1999,Cornish2000,jakethesis} always revealed
that the more time spent near the Feshbach resonance, the greater
the loss. This is still consistent with the basic picture of a
mean-field with some inelastic loss term, albeit a very large one.
Here we report losses occurring on very fast time scales even when
we remain some distance from the resonance, and we see that
shorter and more rapid pulses lead to a greater loss than longer,
slower pulses. This behavior contrasts dramatically with the
conventional picture presented above, and indicates the presence
of novel BEC physics.

To study BEC dynamics we first formed a $^{85}$Rb condensate,
following the procedure given in Ref.~\cite{Cornish2000}. A sample
of $^{85}$Rb atoms in the F=2 m$_{F}$=--2 state was evaporatively
cooled in a cylindrically symmetric magnetic trap
($\nu_{radial}=17.5$ Hz , $\nu_{axial}= 6.8$ Hz). The magnetic
field at completion of evaporation was 162.3~G, corresponding to a
scattering length of 200~$a_0$.  Typically, the cooled sample had
$N_0$=16,500 BEC atoms and fewer than 1000 noncondensed atoms. The
magnetic field was then ramped adiabatically (in 800~ms) to
$\sim$166~G where the scattering length was positive but nearly
zero \cite{Roberts2001a}, and the BEC assumed the shape of the
harmonic trap ground state.

We next applied a short magnetic field pulse (duration $<$ 1~ms)
so that the field briefly approached a value moderately close to
the Feshbach resonance at $\sim$155~G (see Fig.~1).  We used
destructive absorption imaging to look at the number of atoms
remaining in the condensate.  This experiment was repeated with a
variety of differently shaped magnetic field pulses.

We found that the magnetic trap must be turned off and the
condensate spatial size must be significantly larger than our
resolution limit to obtain the most sensitive and accurate
measurements of number \cite{Donley2001}.  To expand the BEC after
the short pulse, we ramped in 5~ms to $\sim$157~G
($a$=1900~$a_0$), and then held at that field for 7~ms.  The
mean-field repulsion during the ramp and hold times decreased the
density by about a factor of 30, then we rapidly turned off the
trap and imaged the cloud.  The density decrease avoided
density-dependent loss that we have observed during the trap
turn-off \cite{Cornish2000, jakethesis}.

Significant number loss from the BEC was observed for pulses
lasting only a few tens of microseconds.  The loss was accompanied
by the generation of a ``burst'' of relatively hot ($\sim$150~nK)
atoms that remained in the trap, as in Ref.~\cite{Donley2001}. The
few thousand burst atoms represented a fairly small fraction of
the total number lost from the BEC and so in this Letter we have
focused only on the number in the condensate remnant.  The burst
will be the subject of future work.

To study the remarkably short time scales for the loss, we
designed a low-inductance, high current auxiliary electromagnetic
coil.  The coil current was supplied by a capacitor bank that was
charged to 580~V, then discharged through the coil at a rate
controlled by a transistor. Our goal was to create a perfect
trapezoidal magnetic field pulse with adjustable but identical
rise and fall times and a hold time during which the field was
constant.  To compensate for observed mutual inductance effects
with other coils in the apparatus, we empirically determined the
auxiliary current pulse shape needed to produce a total magnetic
field pulse that closely approximated our ideal, as shown in
Fig.~1.  The presence of induced currents limited the maximum ramp
speed to $ dB/dt=$1~G/$\mu$s, and the magnetic field uncertainty
was 1 part in 10$^3$ (0.16~G).

\begin{figure}
\includegraphics[bb=150 125 710 469, clip,scale=0.47]{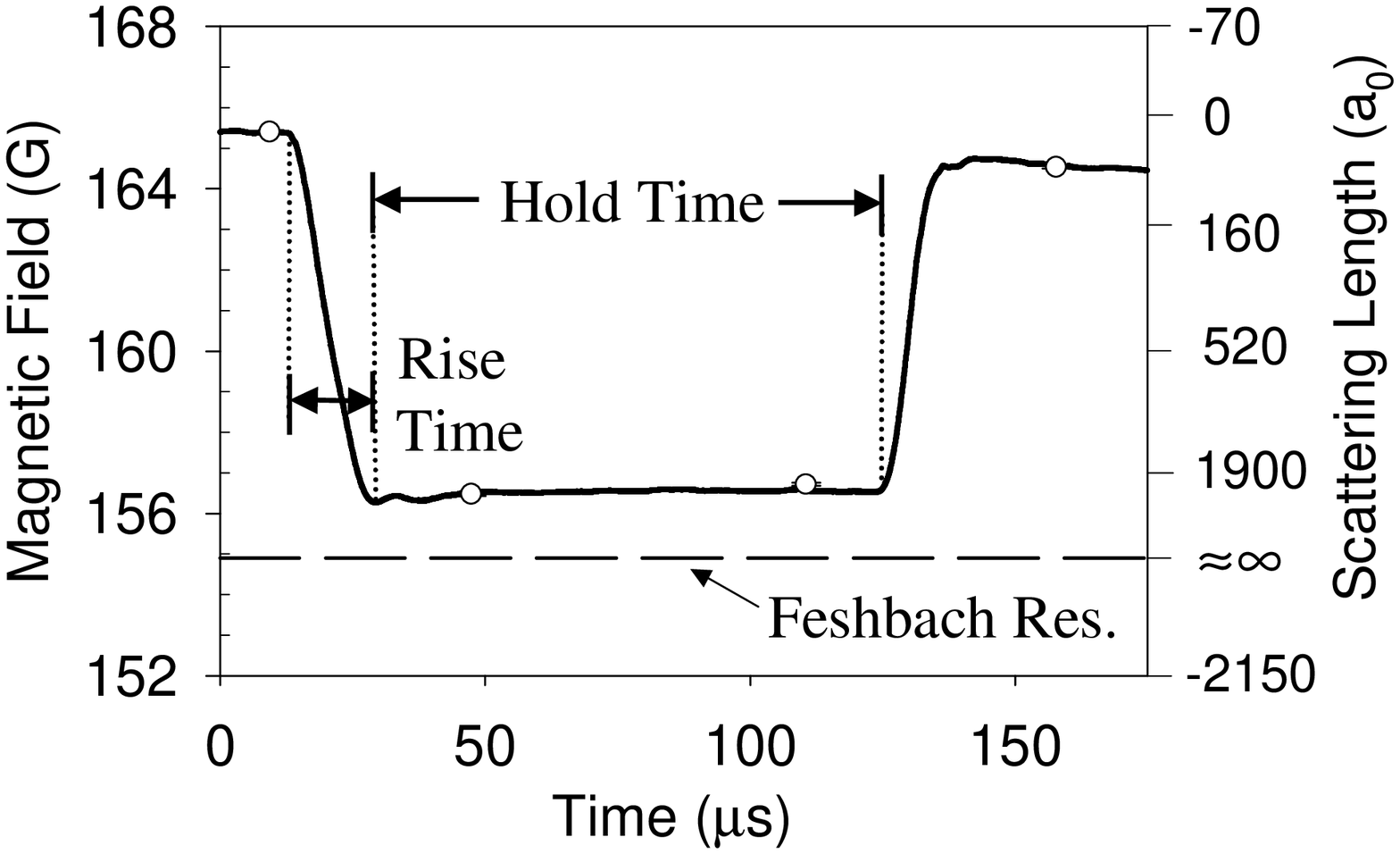}
\caption{The magnetic field vs.~time for a typical auxiliary coil
pulse. The solid line is the field calculated from measurements
with calibrated Hall-effect current sensors on each electromagnet
coil. On the right vertical axis, the corresponding variation in
scattering length is shown.  The open circles are independent
checks of the magnetic field obtained from the resonant frequency
for a 10~$\mu$s RF radiation pulse that drives atoms to the
m$_{F}$=--1 spin state (error bars are smaller than the points).
The dashed line shows the position of the Feshbach resonance,
where the scattering length becomes infinite. The field variation
on the peak of the pulse was typically $\Delta$B $\lesssim$
0.1~G.}
\end{figure}

We first examined the BEC loss for trapezoidal field pulses as a
function of the hold time (see Fig.~2).  Using a linear rise and
fall time of 12.5~$\mu$s, we observed BEC number loss of 10-20$\%$
when the hold time was set to zero (triangular pulse).  The number
of BEC atoms then showed a smooth exponential decrease as hold
time was increased.  Surprisingly, when we reduced the initial BEC
density by more than a factor of 2, the time constant was nearly
unchanged.  For the low[high] density data, the value of the time
constant was over 2[1] orders of magnitude shorter than predicted
by our previous inelastic decay measurements with cold thermal
clouds \cite{Roberts2000}.

\begin{figure}
\includegraphics[bb=144 123 690 469,clip,scale=0.5]{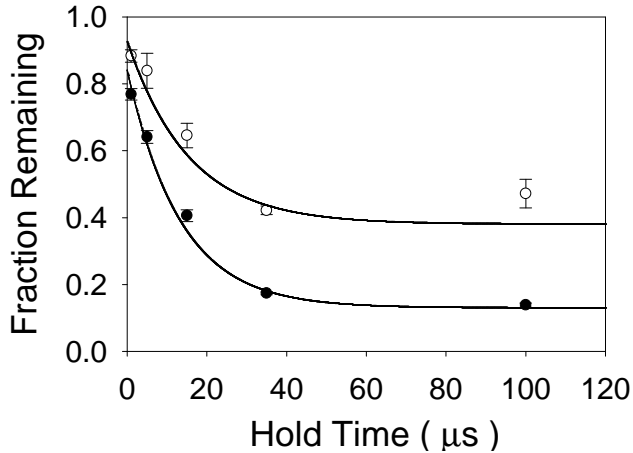}
\caption{Fraction of BEC remaining vs.~pulse hold time.  The pulse
rise/fall time was 12.5~$\mu$s and the magnetic field during the
hold was 156.7~G (2300~$a_0$). Number decay was measured for two
different initial densities: $\langle n\rangle$=1.9x10$^{13}
$~cm$^{-3}$ ($\bullet$) and $\langle n\rangle$=0.7x10$^{13}
$~cm$^{-3}$ ($\circ$).  Fitting the data to exponential functions
(solid lines) gave time constants of 13.2(4)~$\mu$s and
15.4(14)~$\mu$s.  Thus, reducing the density by a factor of 2.6
caused an increase of only 17(11)$\%$ in the time constant.}
\end{figure}

We next measured how the loss depended on the rise/fall time of
the pulse for a variety of pulse amplitudes and hold times
(Figs.~3 and 4). We varied the rise time from 12.5~$\mu$s to
250~$\mu$s and changed the hold time at the pulse peak from
1~$\mu$s to 100~$\mu$s. In addition, the amplitude was varied to
examine fields from 158.0~G ($a$=1100~$a_0$) to 156.0~G
($a$=4000~$a_0$).  Here we list the corresponding scattering
lengths that we have observed by slowly adjusting the magnetic
field, as in Ref.~\cite{Cornish2000}.  The scattering length was
calculated from the equation: $a(B)=a_{bg}(1-\Delta/(B-B_0))$,
where $\Delta$=11.0(4)~G is the width of the resonance,
$B_0$=154.9(4)~G is the resonant magnetic field, and
$a_{bg}$=-450(3)~$a_0$ is the background scattering length
\cite{a_bg}.  For the range of magnetic fields examined here, the
initial value of the diluteness parameter varied from $n a^3$ =
0.01 for $a$=1100~$a_0$ to $n a^3$ = 0.5 for $a$=4000~$a_0$.

\begin{figure}
\includegraphics[bb=150 105 700 485,clip,scale=0.49]{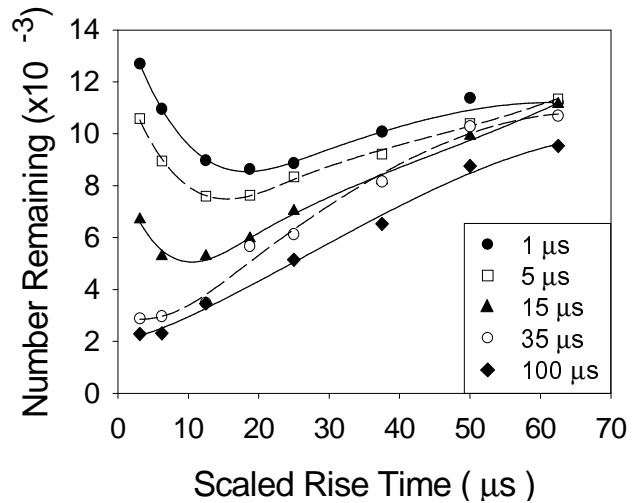}
\caption{Dependence of remnant BEC number on pulse rise/fall time
for various hold times (see legend) with $N_0$=16,500 ($\langle n
\rangle _0 $=1.9x10$^{13}~$cm$^{-3}$).  For the majority of the
data points, the symbol is larger than the statistical error bar
(not shown). The lines are spline fits to guide the eye. The
abscissa was multiplied by a factor of 1/4 to show the time
required to ramp from 75$\%$ to 100$\%$ of the pulse amplitude.
This scaling reflects the observed fact that most of the loss
occurred at fields closest to the Feshbach resonance. The magnetic
field during the hold time was 156.7~G (2300~$a_0$). }

\end{figure}

The number remaining in the BEC after the pulse vs.~pulse rise
time for a variety of different hold times is shown in Fig.~3. For
hold times t$_{hold} \leq$ 15~$\mu$s, there is an initial decrease
in N$_{rem}$ as the rise time increases. Then the slope changes
and fewer atoms are lost for longer rise times. All of the hold
time data display this upward slope over some range, but the range
is largest for the 100~$\mu$s hold time. The increase in remnant
number for longer rise time provides clear evidence that the loss
is not conventional inelastic decay that is characterized by a
rate constant.  It is interesting that the short hold time data
show a distinct minimum in N$_{rem}$ vs.~rise time, which shifts
toward shorter rise times as the hold time is increased.

In Fig.~4 we display the remnant number vs.~ramp time for
triangular magnetic field pulses (1~$\mu$s hold).  The remnant
number is shown for various pulse amplitudes.  For all cases,
N$_{rem}$ is large at the shortest rise time and then decreases
with rise time until it reaches a minimum.  Then longer rise times
cause N$_{rem}$ to increase over a time scale of tens of
microseconds. The rise time that induces maximal loss becomes
longer as the pulses come closer to resonance.

\begin{figure}
\includegraphics[bb=143 100 630 500,clip,scale=0.5]{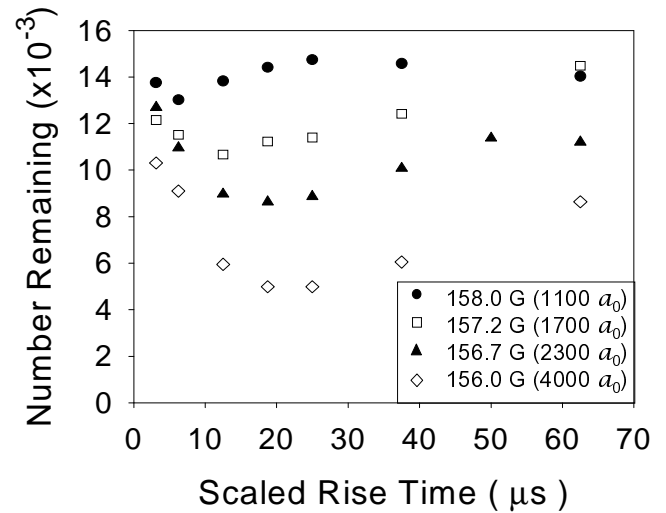}
\caption{Rise time dependence of BEC number for different
triangular pulse amplitudes.  The pulse hold time was fixed at
1~$\mu$s for these data and N$_0$=16,600. Data points are larger
than the error bars.}
\end{figure}

Conventional condensate loss is characterized by a rate constant
for a density-dependent decay process, and thus the loss increases
monotonically with time.  Near a Feshbach resonance the rate
constants have been observed to increase enormously
\cite{Stenger1999,Cornish2000,jakethesis}, but nevertheless, a
longer time spent near the peak, or equivalently a slower ramp
getting over it, resulted in greater loss.  In contrast, we have
measured an increase in the loss when the ramp time is {\it
decreased}, which reveals the existence of previously unexplored
BEC physics.  Of course, the above interpretation of our data
would be modified if the density of the sample was changing due to
the rapid increase in the mean-field interaction, but the
characteristic time for such readjustments in cloud shape is far
longer (of order $1/(2 \nu_{radial})$=29~ms) than the time scales
we considered here.  For example, using a simple analytic model
based on the Gross-Pitaevskii equation \cite{PerezGarcia}, we
calculate that for a 250~$\mu$s ramp to B$_{final}$=156.0~G
(4000~$a_0$), the change in mean-field energy causes the density
to decrease by only 1$\%$ from its initial value. Thus, the
observed dependencies on rise time must reflect microscopic
physics in the BEC and not any macroscopic changes in the shape of
the condensate.

The response of the BEC remnant to magnetic field ramps, observed
in Fig.~4, is qualitatively similar to what one would expect in a
Landau-Zener (L-Z) model \cite{Landau1932,Zener1932} of an avoided
crossing of two states with linear Zeeman shifts, if the second
state were undetectable. In our case, the obvious candidate for
this second state is the Feshbach resonance bound state
corresponding to a diatomic molecule.  With a L-Z avoided
crossing, the behavior when the field approaches and then backs
away from the crossing point with a triangular pulse is
qualitatively similar to the more familiar and analytically
soluble case \cite{L&L} of a linear field ramp that goes from far
above the crossing to far below and then back again. In both
cases, and as seen in Fig.~4, the L-Z model predicts a steep rise
from zero in the transition probability as the length of the ramp
increases from zero (diabatic limit). As the ramp time increases
further, the transition probability reaches a maximum --- where
the time derivative of the relative energy matches the square of
the coupling strength --- and then slowly decreases to zero as the
ramp approaches the adiabatic limit.  We found that the L-Z model
does a rather bad job at reproducing any more quantitative
features of the data however, even when we allowed the coupling
strength and relative magnetic moment to be arbitrary parameters
in our numerical simulations \cite{Rubbmark1981}.

Nevertheless, it seems likely that some atoms are being converted
to another state (possibly molecular) by nonadiabatic mixing,
although the process is more complicated than a simple
Landau-Zener avoided crossing picture.  In future experiments, we
plan to further investigate the time response of the BEC loss
using asymmetric pulses and double pulses with variable spacing.
We will explore the burst production process and attempt to
determine the fate of the lost atoms.

We would like to acknowledge contributions from the entire JILA
BEC/degenerate Fermi gas collaboration and helpful discussions
with L.~Pitaevskii and E.~Cornell.  S.~Thompson acknowledges the
support of the ARO-MURI Fellowship Program.  This work was also
supported by ONR and NSF.

\end{document}